\documentclass[aps,superscriptaddress,
showkeys,twocolumn,amsmath,amssymb,reprint]{revtex4-1}

\usepackage{amsmath}
\usepackage{graphicx}
\usepackage{bm}
\usepackage{color}
\usepackage{upgreek}

\newcommand{\mean}[1]{\langle #1 \rangle}

\begin{document}

\preprint{TES Laser}

\title{Exploring the Photon-Number Distribution of Bimodal Microlasers}

\author{Elisabeth~Schlottmann}
\thanks{schlottmann@\@tu-berlin.de}
\affiliation{Institut f\"ur Festk\"orperphysik, Quantum Devices Group, Technische Universit\"at Berlin, \\
	Hardenbergstra{\ss}e 36, EW 5-3, 10623 Berlin, Germany}
\author{Martin~von~Helversen}
\affiliation{Institut f\"ur Festk\"orperphysik, Quantum Devices Group, Technische Universit\"at Berlin, \\
	Hardenbergstra{\ss}e 36, EW 5-3, 10623 Berlin, Germany}
\author{Heinrich~A.~M.~Leymann}
\affiliation{Max-Planck-Institut f\"ur Physik komplexer Systeme, N\"othnitzer Stra{\ss}e 38, 01187 Dresden, Germany}
\author{Thomas~Lettau}
\affiliation{Otto-von-Guericke-Universit\"at Magdeburg, Universit\"atsplatz 2, 39106 Magdeburg}
\author{Felix~Kr\"uger}
\affiliation{Institut f\"ur Festk\"orperphysik, Quantum Devices Group, Technische Universit\"at Berlin, \\
	Hardenbergstra{\ss}e 36, EW 5-3, 10623 Berlin, Germany}
\author{Marco~Schmidt}
\affiliation{Institut f\"ur Festk\"orperphysik, Quantum Devices Group, Technische Universit\"at Berlin, \\
	Hardenbergstra{\ss}e 36, EW 5-3, 10623 Berlin, Germany}
\affiliation{Physikalisch-Technische Bundesanstalt, Abbestra{\ss}e 2-12, 10587 Berlin, Germany}
\author{Christian~Schneider}
\affiliation{Technische Physik, Universit\"at W\"urzburg, Am Hubland, 97074 W\"urzburg, Germany}
\author{Martin~Kamp}
\affiliation{Technische Physik, Universit\"at W\"urzburg, Am Hubland, 97074 W\"urzburg, Germany}
\author{Sven~H\"ofling}
\affiliation{Technische Physik, Universit\"at W\"urzburg, Am Hubland, 97074 W\"urzburg, Germany}
\affiliation{SUPA, School of Physics and Astronomy, University of St Andrews, St Andrews, KY16 9SS, United Kingdom}
\author{J\"orn~Beyer}
\affiliation{Physikalisch-Technische Bundesanstalt, Abbestra{\ss}e 2-12, 10587 Berlin, Germany}
\author{Jan~Wiersig}
\affiliation{Otto-von-Guericke-Universit\"at Magdeburg, Universit\"atsplatz 2, 39106 Magdeburg}
\author{Stephan~Reitzenstein }
\affiliation{Institut f\"ur Festk\"orperphysik, Quantum Devices Group, Technische Universit\"at Berlin, \\
	Hardenbergstra{\ss}e 36, EW 5-3, 10623 Berlin, Germany}

\date{\today}

\begin{abstract}
A photon-number resolving transition edge sensor (TES) is used to measure the photon-number distribution of two microcavity lasers. The investigated devices are bimodal microlasers with similar emission intensity and photon statistics with respect to the photon auto-correlation. Both high-$\beta$ microlasers show partly thermal and partly coherent emission around the lasing threshold. For higher pump powers, the strong mode of microlaser \textbf{A} emits Poissonian distributed photons while the emission of the weak mode is thermal. In contrast, laser \textbf{B} shows a bistability resulting in overlayed thermal and Poissonian distributions.  While a standard Hanbury Brown and Twiss experiment cannot distinguish between simple thermal emission of laser \textbf{A} and the mode switching of laser \textbf{B}, TESs allow us to measure the photon-number distribution which provides important insight into the underlying emission processes. Indeed, our experimental data and its theoretical description by a master equation approach show that TESs are capable of revealing subtle effects like mode switching of bimodal microlasers. As such our studies clearly demonstrate the huge benefit and importance of investigating nanophotonic devices via photon-number resolving sensors.
\end{abstract}

\keywords{Microlaser, Cavity Quantum Electrodynamics, Photon-number distribution, Photon statistics, Photon-number resolving detectors} 

\maketitle
\textbf{Introduction.---}Microlasers are of enormous interest for both fundamental research of cavity enhanced nanophotonic devices and their future applications due to their small size, high speed and low energy consumption~\cite{Khurgin2012}. Popular microlaser concepts are based on photonic crystal cavities~\cite{Nomura2010}, plasmonic resonators~\cite{Oulton2009} or micropillar cavities~\cite{Reitzenstein2006, Lermer2013}. These resonator structures have small mode-volumes in common, which result in enhanced light-matter coupling. As a consequence, the associated spontaneous emission factor $\beta$ is strongly enlarged so that the ultimate limit of thresholdless lasing can be approached~\cite{Bjork1994}.

In devices with high $\beta$-factors, analyzing the input-output characteristics is not sufficient to prove lasing operation due to the lack of a significant nonlinearity at the threshold. Furthermore, optical injection, superradiance, mode competition and saturation of the low dimensional gain medium can also lead to deviations from the standard behaviour~\cite{Schlottmann2016, Siegman1986, Rice1994, Leymann2013, Jahnke2016, Kreinberg2017}. A well established method to analyse the statistics of the emitted light is based on the  Hanbury Brown and Twiss (HBT) configuration~\cite{Hanbury-Brown1956}, which essentially measures the time correlation of photon pairs to determine the second order auto-correlation function $g^{(2)}(\tau)$. Studying the photon statistics has become an important tool to characterize microlasers as it reveals the transition from predominantly spontaneous emission towards stimulated emission at threshold by a change of $g^{(2)}(0)$ from 2 to 1~\cite{Ulrich2007}. Interestingly, in bimodal lasers additional effects like gain competition~\cite{Leymann2013, Redlich2016} and dissipative coupling~\cite{Fanaei2016} occur which are difficult to identify by a HBT measurement alone. We show that a full understanding of the processes involved in the emission of such nanophotonic devices requires not only information quantified in $g^{(2)}(0)$ but also knowledge of the photon-number distribution.

\begin{figure}[htb]
	\center
	\includegraphics[width=1\columnwidth]{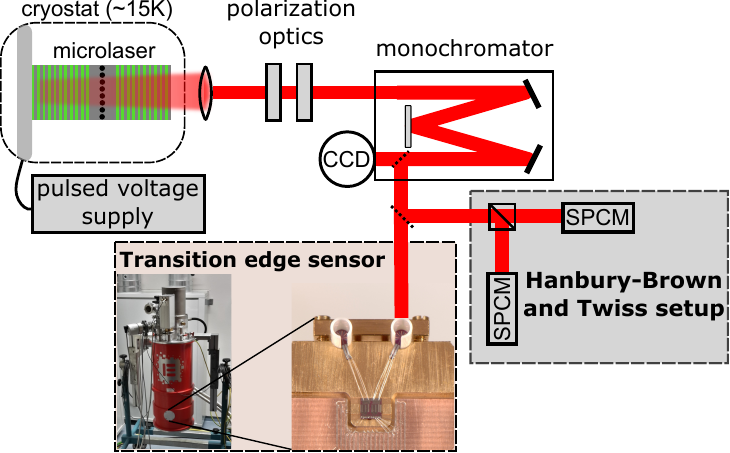}
	\caption{Sketch of the experimental setup: The microlaser sample is operated in a He-flow cryostat at 15 K and is excited by a pulsed electrical voltage supply. The emitted light is analyzed by a spectrometer, the TES or alternatively by a standard HBT setup.}
	\label{fig1}
\end{figure}

This challenge can be addressed by using photon-number resolving detectors capable of determining the photon-number distribution of the emission. Unfortunately, standard single-photon sensitive detectors based on avalanche photo-diodes are not capable of determing the number of impinging photons. This is different for another class of highly efficient detectors - namely transition edge sensors (TES, see Fig.~\ref{fig1}). Such detectors have usually high quantum efficiency in excess of 90\% over a large range of wavelengths~\cite{Lita2008}, and can be used as photon-number resolving detectors because of their calorimetric operation principle~\cite{Cabrera1998, Irwin2005}. Interestingly, despite of the huge benefit of being able to experimentally acess the photon-number distribution of ultra-low light-level emitters~\cite{Miller2003}, TESs have not been applied for the in-depth analysis of nanophotonic devices.  

In this letter, we apply a TES to measure the photon-number distribution of two microlasers with two orthogonally polarized modes. This allows us to obtain deeper insight into the emission properties which is hardly possible using standard characterization tools such as a HBT configuration. Our work also highlights the enormous - and so far uncovered - potential of TESs as an important measurement concept in the application of microlasers and in the wide field of nanophotonics. To illustrate this potential we select two bimodal microlasers with, at first sight, very similar emission features. For the first laser \textbf{A} the emission of both modes is in a transient state from thermal to coherent light around the laser threshold. While for high pump rates the stronger mode emits pure coherent light, the weaker mode is in the thermal regime. The second laser \textbf{B} has similar input-output characteristics and $g^{(2)}(0)$-values. Excitingly, for this laser gain competition between the two emission modes leads to mode switching and an associated double-peaked photon-number distribution. The latter can only be revealed by the TES technique and is best described by an overlay of thermal and Poissonian statistics.

\textbf{Theoretical Methods.---}To calculate the full photon statistics $P_{n}$ of emission from the microlaser we solve a master equation for the diagonal elements of the density matrix $\rho^{\bf n}_N$ giving the probabilities to find the system in a state with photon numbers ${\bf n}=(n_{\text{w}},n_{\text{s}})$ in the
weak- and in the strong mode of the laser and $N$ excited emitters. The master equation is a multi-mode generalization of the equation used in~\cite{Rice1994} and is based on a statistical birth-death model including all relevant processes of a multi mode laser on a phenomenological level. This model has been applied successfully to bimodal microcavity lasers before, to address the origin of super thermal intensity fluctuations~\cite{Leymann2013} and to investigate the connection between non-equilibrium Bose-Einstein condensation~\cite{Vorberg2013} and pump power driven switching of the lasing mode~\cite{Leymann2017}. To describe the detection of photons emitted by the microlaser with the TES, a detection model introduced in~\cite{Lee1993}, is used. The pulsed excitation and detection applied in the present work is theoretically described by two steps: First the steady state of the laser system is found for a pump rate corresponding to the pump area. Second this steady state decays via the leaky cavity and the leaked and detected photons are counted (for further details see Appendix \ref{app:Theory}).\\

\textbf{Sample Technology and Experimental Setup.---}
The gain medium of the used microlasers is composed of a single layer of In$_{0.3}$Ga$_{0.7}$As quantum dots with a density of $5\cdot 10^9 /$cm$^2$. The active layer is embedded into the central one-$\lambda$ GaAs cavity which is sandwiched between an upper (lower) distributed Bragg reflector consisting of 26 (30) mirror pairs that are based on $\lambda/4$-thick layers of GaAs and AlAs. Micropillars of 4 $\upmu$m diameter are produced via electron beam lithography and plasma etching. The sample is planarized with benzocyclobutene and individual micropillars are electrically contacted with circular gold contacts. The Q-factor of the electrically contacted micropillars is about 20.000. Details on the sample fabrication are explained in Ref.~\cite{Boeckler2008}.

The microlaser sample is placed in a continuous flow He-cryostat and cooled down to a temperature of $T$\;=\;15\;K (c.f. Fig.~1). It is pumped by an electrical pulse generator with variable pulse length (0.5-10~ns) and pulse amplitude up to 5.1\;V$_{\text{AC-Bias}}$ and a repetition frequency of 10\;kHz. A bias voltage V$_{\text{Bias}}$=V$_{\text{DC-Bias}}$+V$_{\text{AC-Bias}}$ with V$_{\text{DC-Bias}}$=1.5\;V is applied. For laser \textbf{A} a pulse length of $\tau_{P}$\;=\;2\;ns and for laser \textbf{B} a pulse length of $\tau_{P}$\;=\;1.5\;ns is chosen. A microscope objective collects the emission. Polarization optics are used to separate the two orthogonal modes, and their emission is spectrally resolved by a spectrometer with a resolution of 30\;$\upmu$eV. Finally, the signal is analyzed with a TES or alternatively by a HBT setup.

The TES acts as a highly sensitive calorimeter to detect the small energy input from an absorbed photon pulse. The temperature change is measured with a sensitive thermometer which is simultaneously the absorber. By voltage biasing, the TES heats up within the superconducting phase transition and is stabilized by negative electro-thermal feedback~\cite{Lita2008} so that the absorption of a photon pulse results ultimately in a current redistribution. The current change is measured via an inductively coupled two-stage dc-superconducting quantum interference device (SQUID)~\cite{Drung2007}. The TES/SQUID detector unit is fiber-coupled and mounted on the cold stage of an adiabatic demagnetization refrigerator, which is stabilized at 130\;mK. From analyzing many pulses, a histogram of the photon-number distribution can be extracted. The detection efficiency of the TES is determined to be 87~\%.\\

\textbf{Experimental results.---}
\begin{figure}[htb]
	\center
	\includegraphics[width=1\columnwidth]{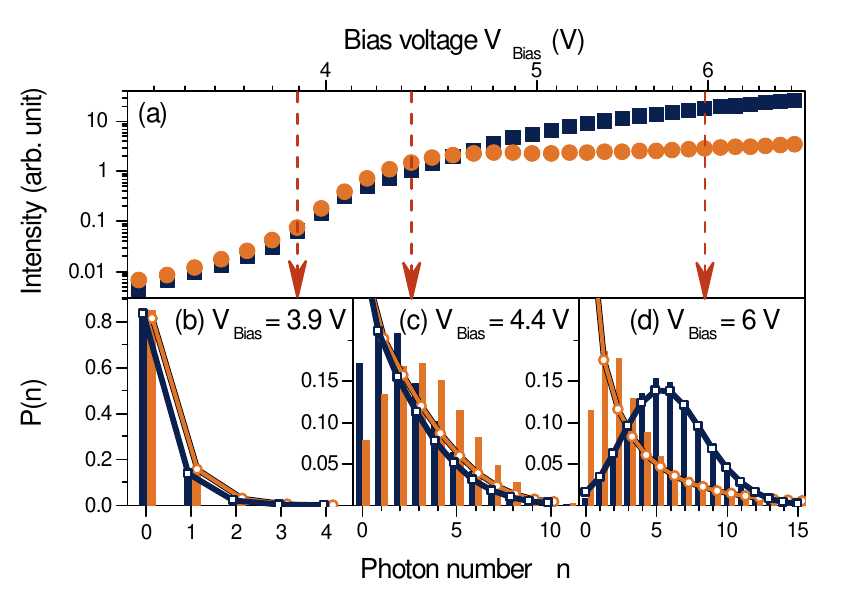}
	\caption{(a) Intensity-bias voltage characteristic of laser~\textbf{A}. The strong mode (blue squares) shows an s-shaped behavior while the weak mode (orange circles) saturates in intensity above the threshold. (b)-(d): Photon statistics at three bias voltages, indicated by the red arrows. The bars correspond to the statistics measured with the TES, the dots, connected by a line, correspond to the theory. (b) For low bias voltage ($V_{\text{Bias}}=3.9$ V) both modes possess a Poissonian distribution. (c) Above threshold ($V_{\text{Bias}} = 4.4$ V), the photon statistics exhibits a transient distribution, which is partly thermal and partly Poissonian. (d) For high voltage ($V_{\text{Bias}} = 6.0$ V), the weak mode shows a thermal distribution whereas the strong mode exhibits a Poissonian distribution.}
	\label{fig2}
\end{figure}
The investigated micropillar lasers have two nearly degenerated fundamental modes that can, however, be separated by their orthogonal polarization. Both fundamental modes couple to the common gain medium and experience gain competition, while higher-order modes can be neglected. The intensity-bias voltage dependence of laser \textbf{A} [Fig.~2 (a)] reveals the typical behavior: At first, both modes increase super linearly at the threshold, then at higher excitation gain competition leads to a decrease in intensity in the weak mode (orange circles) and a further increase in the strong mode (blue squares)~\cite{Leymann2013}. 

Figure~2 (b)-(d) depict the photon-number distribution for three voltages. For low voltage pulses, both modes have a Poissonian distribution. The microlaser is expected to emit thermal light, but since the coherence time is shorter than the pulse length $\tau_{coh}\ll \tau_{P}$, the real character is not accessible in this regime since thermal bunching arises on a scale of the coherence time~\cite{Ulrich2007}. Therefore, a longer pulse averages over many bunching events and a Poissonian distribution is measured~\cite{Loudon1983}. The coherence time at the bias voltage of 3.9 V can be estimated from the linewidth as $\tau_{coh}\sim170$\;ps~\cite{Ates2007}. The theoretical calculations (dots connected by a line) which will be detailed below do not suffer from coherence time limitations and reproduce a thermal distribution almost perfectly. For these low photon numbers the two distributions are almost indistinguishable by the eye.

Above the threshold at $V_{\text{Bias}} = 4.4$ V both modes are in a transient state and the photon-number distribution is partly thermal and partly Poissionian~\cite{Arecchi1966}. In this mixed photon-number distribution the Poissionian part, which indicates the emission of coherent light, is recognizable by the enhanced contribution of higher photon numbers. Our theory describes the same behavior, however, without coherence time limitation, it predicts a higher probability for zero-photon events if compared to the experimental data.

For a high bias voltage of 6.0~V, the photon-number distribution of the weak and the strong mode differ considerably. Whereas the strong mode emits pure coherent light, indicated by Poissonian statistics, emission of the weak mode has thermal properties. The experimental photon-number distribution of the strong mode is in very good agreement with the theory. Since the coherence time ($\tau_{coh,\text{w}}\sim 530$\;ps) is shorter than the pulse length ($\tau_{p}\sim 2$\;ns), a pure thermal distribution cannot be measured for the weak mode. This explains again the deviation between theory and experiment noticeable at low photon numbers $\leq$\;5.

\begin{figure}[htb]
	\center
	\includegraphics[width=1\columnwidth]{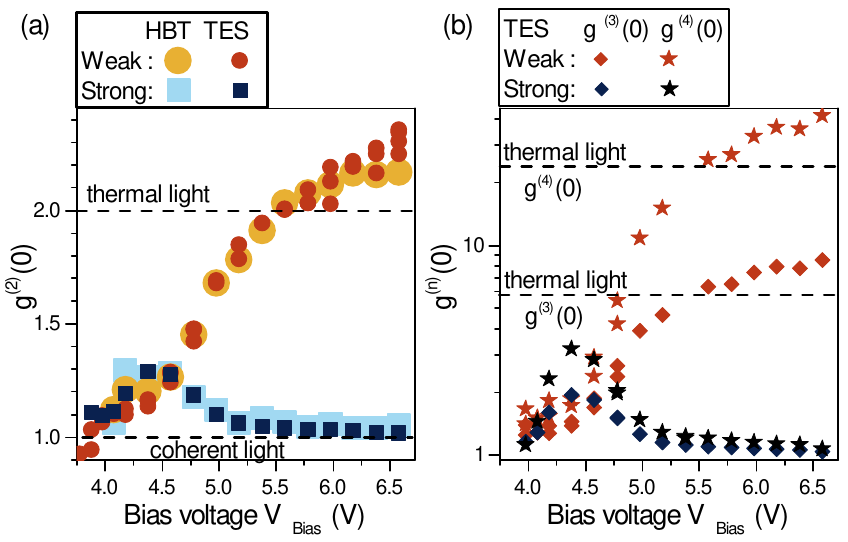}
	\caption{Laser \textbf{A}: (a) The second order auto-correlation function at zero-time delay $g^{(2)}(0)$ determined by the TES (small darker symbols) is in very good agreement with the one determined by the HBT (big brighter symbols). The strong mode shows a transition from thermal to coherent emission. The weak mode increases in $g^{(2)}(0)$ to values slightly above the thermal limit. (b) Third and fourth order of the auto-correlation function $g^{(n)}(0)$ from the same TES measurements exhibit a behavior analogue to the one of $g^{(2)}(0)$.}
	\label{fig3}
\end{figure}

Interestingly, while standard HBT measurements provide only information about the second-order autocorrelation function, all moments and hence all orders of the auto-correlation function at zero time delay $g^{(k)}(0)$, can be calculated from the experimentally determined photon-number distribution $P_{n}$~\cite{Migdall2013}:
\begin{eqnarray}
g^{(k)}(0)=\frac{\sum_{n}\prod_{i=0}^{k-1}(n-i)\cdot P_n}{\langle n\rangle^k}\ .
\label{eq:gk}
\end{eqnarray}
To determine the second order auto-correlation function $g^{(2)}(0)$ only the mean photon number $\langle n\rangle=\sum_{n}n\, P_{n}$ and the variance Var$(n)=\langle (n-\langle n\rangle)^2 \rangle$ are required:
\begin{eqnarray}
g^{(2)}(0)=1+\frac{\text{Var}(n) - \langle n\rangle }{\langle n\rangle^2}.
\label{eq:g2}
\end{eqnarray}
In Fig. 3 (a) the $g^{(2)}(0)$-values of laser \textbf{A} for varied pulse voltage are presented. The data calculated from the TES measurements is close to perfect agreement with the corresponding HBT data. For low voltage, the thermal emission with an expected $g^{(2)}(0)$\,=\,2 is, as already discussed, not resolvable and a $g^{(2)}(0)$\,=\,1 is measured. In the transition region, an increase up to 1.3 is visible. This represents the transition from thermal emission to lasing operation with the simultaneous increase of the coherence time~\cite{Strauf2006,Ulrich2007}. The auto-correlation of the strong mode decreases to $g^{(2)}(0)$\,=\,1 for higher voltage, indicating coherent emission. For the weak mode, the auto-correlation increases first and then stabilizes at $g^{(2)}(0)$ slightly above 2. This behavior, i.e. $g^{(2)}(0) > 2$ is an indication for thermal emission with minor contributions of other effects like superradiance that are beyond the scope of this article. The accordance of both techniques proves the accuracy of the determined $g^{(2)}(0)$. 

The third and fourth order of the auto-correlation function [see Eq.~\ref{eq:gk}] obtained from the TES data are exemplarily depicted in Fig.~3 (b). The different orders of $g^{(k)}(0)$ follow the same trend as $g^{(2)}(0)$, but reach higher values. The dashed lines indicate the respective thermal limits $k!$. Being able to address higher-order photon autocorrelation functions to e.g. better understand the threshold behaviour of micro lasers \cite{Leymann2014} is another advantage of the TES technique. Indeed, higher order auto-correlations cannot be accessed by standard HBT experiments and up till now only elaborate streak-camera measurements allowed to access the auto-correlation function up to fourth order~\cite{Assmann2009, Wiersig2009}.

\begin{figure}[htb]
	\center
	\includegraphics[width=1\columnwidth]{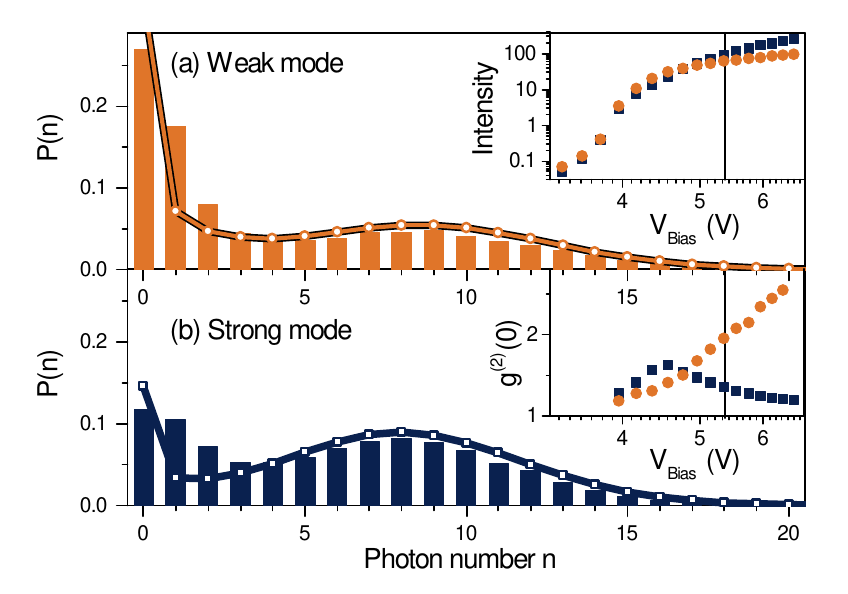}
	\caption{Laser \textbf{B}: The photon-number distributions of the weak (a) and strong (b) mode  at$V_{\text{Bias}}=5.4$ V can best be described by an overlay of a thermal distribution with a low mean photon number $\langle n  \rangle$ and a Poissonian distribution with high $\langle n \rangle$. The input-output characteristic (inset (a)) and $g^{(2)}(0)$-values (inset (b)) of laser \textbf{B} are similar to laser \textbf{A} (cf. Fig.~\ref{fig2} and Fig.~\ref{fig3}).}
	\label{fig4}
\end{figure} 

To highlight the importance of investigating microlasers with a TES, a second laser \textbf{B} with almost identical input-output and auto-correlation characteristics (see insets of Fig.~4) is investigated. Analyzing its full photon statistics, only accessible with a TES, we see substantial differences between laser \textbf{A} and laser \textbf{B}. Compared to laser \textbf{A} both, the weak and the strong mode, show a behaviour with an emission being composed of a thermal distribution with a low mean photon number $\langle n \rangle$ and a Poissonian distribution with a large $\langle n \rangle$. In striking contrast to the statistics of the laser \textbf{A}, for laser \textbf{B} the zero photon state is the most likely one for both the strong and the weak mode. The difference between the weak and the strong mode results in the fact that the emission statistics of the former mode is dominated by the thermal part, whereas the strong mode is dominated by the Poissonian part. This behavior can be explained as follows: Both modes are potential lasing modes where carrier fluctuations largely influence the switch-on process. 

For every electrical pulse, potentially each of both modes could reach the lasing regime while the other mode stays in the thermal regime. In the presented case, the analysis of the experimental photon-number distributions yields that in $\sim$\,75\,\% of the pulses, the strong mode is in the lasing regime and emits coherent light while the weak mode radiates thermally. In the other $\sim$\,25\,\% of the pulses the weak mode is lasing and the strong mode is not. This manner is comparable to spontaneous switching under continous wave excitation~\cite{Redlich2016, Leymann2013}. Also the theoretical description reproduces this behavior well. In the master equation the spontaneous transition between the modes is effectively reduced (compared to laser \textbf{A}) due to stronger modal interactions and carrier population oscillations~\cite{Marconi2016}, thus trapping the weak mode close to the zero photon state and giving rise to the bistable behavior~\cite{Leymann2017}. \\

\textbf{Conclusion.---}
We have demonstrated that TESs are powerful detectors to investigate the photon statistics of microscopic laser devices. Where former HBT experiments are only able to detect intensity fluctuations quantified in $g^{(2)}(0)$ regardless of their origin, the TES gives direct access to the photon-number distribution and enables the differentiation between various effects. Determining the full photon statistics via TES detectors has high potential to become a powerful characterization method to reveal and understand the physics of nanophotonic devices at the quantum level. It will be of particular importance for the further development of microcavities towards applications which benefit from a tunable and controllable photon statistics of emission. 

\begin{acknowledgments}
	The research leading to these results has received funding from the European Research Council under the European Union's Seventh Framework ERC Grant Agreement No. 615613, within the EURAMET joint research project MIQC2 from the European Union's Horizon 2020 Research and Innovation Programme and the EMPIR Participating States and from the German Research Foundation within the project RE2974/10-1. The authors thank the State of Bavaria for financial support and A. E. Lita and S. W. Nam for providing the TES detector chips.
\end{acknowledgments}

\appendix
\section{Details of the Theoretical methods}
\label{app:Theory}
To describe the measurement theoretically, we divide the process in two subprocesses: (i) the excitation of the laser device by the pump pulse and (ii) the subsequent detection of the emitted cavity photons. The first subprocess is modeled by the steady state of the master equation Eq.~(\ref{eq:master}), which is determined by solving the linear equation $\frac{d}{dt}\rho^{\bf n}_{N}=0$ (see \ref{app:master}). This steady state is then modified according to~\cite{Lee1993} (see \ref{app:detect}).

\subsection{Master equation}
\label{app:master}

The utilized master equation

\begin{align}
	\frac{d}{dt}\rho^{\bf n}_{N}
	=&P\left[\rho^{\bf n}_{N-1}-\rho^{\bf n}_N\right]
	-\tau^{-1}[N\rho^{\bf n}_N-(N+1)\rho^{\bf n}_{N+1}]
	\nonumber \\ &
	-\sum_i g_{i}[N(n_{i}+1)\rho^{\bf n}_N
	-(N+1)n_{i}\rho^{{\bf n}-{\bf e_i}}_{N+1}]
	\nonumber\\&
	-\sum_i \ell_{i}[n_{i}\rho^{\bf n}_N
	-(n_{i}+1)\rho^{{\bf n}+{\bf e_i}}_N]
	\nonumber\\&
	-\sum_{i,j}R_{i\to j}\big[n_{i}
	(n_{j}+s)\rho^{\bf n}_N
	\nonumber\\&
	-(n_{i}+1)(n_{j}-1+s)\rho^{{\bf n}+{\bf e_i}-{\bf e_j}}_{N}\big].
	\label{eq:master}
\end{align}

is based on a phenomenological model that takes all of the relevant processes of the microcavity laser into account.
Here $P$ is the pump rate, $\tau^{-1}$ the rate of spontaneous emission into non-lasing modes, $g_{i}$ is the rate of emission into the lasing mode~$i$, $\ell_{i}$ is the loss rates of photons from cavity $i$ and $R_{i\to j}$ is the transition rate of the cavity photons from mode $i$ to mode $j$. $s$ is the factor quantifying how strong the gain medium induced mode interaction effectively reduces the spontaneous emission between the modes.
The solution of Eq.~(\ref{eq:master}) can be interpreted as the diagonal elements of the density matrix $\langle {\bf n}, N|\rho|{\bf n},N \rangle=\rho^{\bf n}_{N}$, giving the probability to find the system with $N$ excited emitters and ${\bf n}=(n_{\mathrm{w}},n_{\mathrm{s}})$ photons in the weak and strong mode respectively. By tracing over the emitters and one of the modes one can obtain for example the distribution of the weak mode $P_{n_{\mathrm{w}}}=\sum_{N,n_\mathrm{s}}\rho^{\bf n}_{N}$. The parameters for the theory are given in Tab.~\ref{tab:1}.
\begin{table}[htb]
	\centering
	\caption{Simulation parameters used in Figures 2 and 4}
	\begin{ruledtabular}
	\begin{tabular}{cll}
		Parameter & Fig.~2 & Fig.~4  \\
		\colrule
		s         &    1   & 0\\
		\colrule
		in units of $\tau$ & & \\
		$l_1$ & 0.1 & 0.1 \\
		$l_2$ & 0.105 &  0.105 \\
		$g_1$ & 0.14 & 0.12 \\
		$g_2$ & 0.12 & 0.1 \\
		$R_{21}$ & 0.003 & 0.004 \\
		$R_{12}$ & 0.00325 & 0.00425 \\
        \colrule
        in units of $P_{thr}$ & & \\
		$P_b$ & 0.24 & \\
		$P_c$ & 1.3 &   \\
		$P_d$ & 8.9 &  \\
		$P$ &  & 4.6 
		\label{tab:1}
	\end{tabular}
\end{ruledtabular}
\end{table}

\subsection{Detection model}
\label{app:detect}
Since the master equation models the inside of the cavity, it is necessary to study the change of the statistics with respect to the leakage of photons out of the cavity $\ell_{i}$ and the non-ideal setup, with an efficiency denoted by $\xi$. Assuming that the leakage of the cavity is the relevant process, i.~e., the pump pulse has already subsided and the rate of the intermode kinetics is comparable small, the influence of the detection for a single mode distribution can be modeled as

\begin{align}
	P^{\text{out}}_{m}(t_1, t_2) = \sum_{n_i=m} &P_{n_i} \left({ {n_i}\atop{m} }\right) \left( 1 - \xi e^{-\ell_i t_1} + \xi e^{-\ell_i t_2} \right)^{n_i - m}\nonumber\\
	&\times\left(\xi e^{-\ell_{i} t_{1}} - \xi e^{-\ell_{i} t_{2}}\right)^m ,
	\label{AppEq:pout}
\end{align}

where $P_{n_i}$ is the single mode distribution (see \ref{app:master}), $P^{\text{out}}_{m}$ is the detected distribution and $t_1$ and $t_2$ are the times at which the measurement begins and ends, respectively~\cite{Lee1993}. Although this transformation shifts the whole statistics to a lower mean number, it does not alter the photon auto-correlation $g^{(2)}(0)$.
To proof this we define $\zeta = \xi e^{-\ell_{i} t_{1}} - \xi e^{-\ell_{i} t_{2}}$ and find that $ \mean{n}_{{\text{out}}}$ and $ \mean{n^2}_{{\text{out}}}$ can be expressed by  $\zeta$ and the expectation values inside the cavity by

\begin{align}
	\mean{n}_{{\text{out}}} &= \zeta \mean{n},\nonumber\\
	\mean{n^2}_{{\text{out}}} &= \zeta^2\mean{n^2} + (\zeta - \zeta^2)\mean{n}.
	\label{AppEq:mean}
\end{align}

This follows from Eq.~(\ref{AppEq:pout}) by changing the order of summation and using the knowledge of the mean and the variance of the binomial distribution.

Relations (\ref{AppEq:mean}) can be inserted in Eq.~(\ref{eq:g2}) and it follows that the transformation $P^{\text{out}}_{m}(t_1, t_2)$ does not change $g^{(2)}(0)$. The setup efficiency is estimated to be $\xi=0.1$. Since the measurement lasts much longer than the cavity decay time, we set $t_2\rightarrow\infty$ and $t_1=0$, since the initial state for the detection model is the steady state of Eq.~(\ref{eq:master}).

\end{document}